\documentclass[runningheads]{llncs}
\usepackage{graphicx}
\begin{document}

\title{Gap-Townes solitons and delocalizing transitions of multidimensional Bose-Einstein condensates in
 optical lattices}

\titlerunning{Gap-Townes solitons and delocalizing transitions  of multidimensional BEC}

\author{Mario Salerno$^{1}$ \thanks{E-mail: salerno@sa.infn.it}, F. Kh. Abdullaev$^{2}$, and B. B. Baizakov$^{2}$}
\institute{ $^1$ Dipartimento di Fisica \textquotedblleft E.R.
Caianiello" and Consorzio Nazionale Interuniversitario per le
Scienze Fisiche della Materia (CNISM), Universit\'{a} di Salerno,
I-84081 Baronissi (SA), Italy \\
$^2$ Physical-Technical Institute of the Uzbek Academy of
Sciences, \\ 100084, Tashkent, Uzbekistan}

\maketitle

\begin{abstract}
We show the existence of gap-Townes solitons for the
multidimensional Gross-Pitaeviskii equation with attractive
interactions and in two- and three-dimensional optical lattices.
In absence of the periodic potential the solution reduces to the
known Townes solitons of the multi-dimensional nonlinear
Schr\"odinger equation, sharing with these the propriety of being
unstable against small norm (number of atoms) variations. We show
that in the presence of the optical lattice the solution separates
stable localized solutions (gap-solitons) from decaying ones,
characterizing the delocalizing transition occurring in the
multidimensional case. The link between these higher dimensional
solutions and the ones of one dimensional nonlinear Schr\"odinger
equation with higher order nonlinearities is also discussed.
\end{abstract}

\section{Introduction}

One interesting phenomenon occurring in ultracold atomic gases
trapped in periodic potentials is the possibility to localize
matter in states which can stay for a long time due to an
interplay between nonlinearity, dispersion and periodicity. Such
states (also called gap-solitons)  have been observed in
Bose-Einstein condensates (BEC) and in arrays of nonlinear optical
waveguides \cite{Burger,Khaykovich,Strecker,Eiermann}. For
attractive atomic interactions in BEC and in absence of a periodic
potential, stable localized solutions are possible only in a
one-dimensional (1D) setting since in two (2D) and three (3D)
dimensions the phenomenon of collapse appears \cite{Sulem}. More
precisely, one observes that when the number of atoms exceeds a
critical threshold, the solution collapses in a finite time
(blow-up) while for number of atoms below the critical threshold
there is an irreversible decay of the state into background
radiation. For the higher dimensional nonlinear Sch\"odinger (NLS)
equation, however, it is known that there exists an unstable
localized solution, the so called Townes soliton \cite{townes},
which separates decaying solutions from collapsing ones. Townes
soliton, however,  exists only for a single value of the number of
atoms, being unstable against fluctuations around it (for slightly
overcritical or undercritical number of atoms the solution
collapses or decays, respectively). The situation is drastically
changed in presence of an  optical lattice (OL). To this regard,
it has been shown that stable 2D and 3D solitons can exists in OLs
both in BEC and nonlinear optics contexts
\cite{BKS,Efremidis02,OK03,BMS03,Lederer05}. Moreover, it is known
that while the periodic potential can only marginally shift the
critical value for collapse, it can substantially move the
delocalizing transition curve, thereby increasing the soliton
existence range in parameter space  from a single point to a whole
interval \cite{BS04}. The typical situation with 2D and 3D BEC
solitons in OLs is therefore the following: in the parameter space
the stable localized solutions are confined from above by the
collapse curve and from below by the delocalizing transition
curve, thus,  in contrast with the one dimensional case where
there are no limits for the existence of localized states, strict
limitations for soliton existence appear in multidimensional
cases. From this point of view it is clear that for possible
experimental observation of multidimensional BEC solitons the
parameter design becomes very important. Since the collapse curve
is only marginally affected by the periodic potential, to enlarge
existence ranges of solitons it is of interest to give a full
characterization of  the delocalizing curve in parameter space.

The aim of this paper is just devoted to this, i.e. we
characterize  2D and 3D delocalizing curves of gap-solitons in
terms of an unstable solution of the multidimensional
Gross-Pitaeviskii equation (GPE), which we call gap-Townes
soliton. This solution can be viewed as a separatrix (it separates
gap soliton states from extended (Bloch) states) and reduces to
the known Townes soliton when the strength of the OL goes to zero.
Similar solutions were found also for the 1D NLS equation with
higher order nonlinearities  in  \cite{AS05} , where they were
called gap-Townes solitons, and in \cite{KP}  where they were
termed Townes solitons. Conditions for the occurrence of the
delocalizing transition phenomenon of one-dimensional localized
modes of several nonlinear continuous periodic and discrete
systems of the nonlinear Schr\"odinger type were also recently
discussed in \cite{CBKS08}. For the periodic multidimensional  GPE
the delocalizing curve has been characterized in \cite{BS04} as
the critical threshold for the existence of one bound state in an
effective potential. The characterization given here, however, is
more general since it is valid also for 1D NLS with higher order
nonlinearities. To this regard we remark that in absence of
confining potential the 2D and 3D GPE behaves similarly to the 1D
NLS with quintic and septic nonlinearities, respectively. The
interplay between dimensionality and nonlinearity has been used to
investigate collapse in lower dimensional NLS on the basis of pure
dimensional arguments. In particular, the critical condition for
collapse has been characterized as $D(n-1)-4=0$, where $n$ is the
order of the nonlinearity in the equation and $D$ is the
dimensionality of the system \cite{Berge}. In the following we
take advantage of this interplay to construct approximate
gap-Townes soliton solutions of the GPE with multidimensional
separable OLs, in terms of products of exact gap-Townes solutions
of the 1D NLS with higher nonlinearities. Remarkably, we find
that, except for strengths of the optical lattices very small, our
approach produces very accurate gap-Townes solutions of
multidimensional GPE with OL, thus giving  an evident
computational advantages. The results obtained in this paper can
be seen as a generalization of the existence of gap-Townes
solitons in the quintic NLS discussed in \cite{AS05,KP} to the
case of the multidimensional Gross-Pitaeviskii equation.

We finally remark that the obtained results can also be applicable
for photonic lattices with Kerr type of optical nonlinearity where
the existence of a critical threshold for the lattice solitons has
been observed \cite{Efremidis03}.

The paper is organized as follows. In section II we introduce the
model equations and discuss the link between multidimensional GPE
with a separable trapping potential and the corresponding 1D NLS
equation with higher order nonlinearity. We use a self consistent
approach to approximate gap-Townes solitons of the GPE with
products of exact gap-Townes soliton of the corresponding 1D NLS
equation with higher order nonlinearity. In Section III we discuss
the existence of localized solutions in the multidimensional GPE
with OL by means by a variational approach and compare 2D and 3D
results with those obtained from the VA applied to the quintic and
septic NLS, respectively. In section IV we perform a numerical
investigation of the existence (delocalizing) threshold for
gap-Townes solitons of the 2D and 3D GPE. Finally, in the last
section we briefly summarize our main results.

\section{Model equations and existence of gap-Townes solitons}

Let us  consider the following Gross-Pitaevskii equation in
d-dimensions ($d=1,2,3,$) as a model for a BEC in an optical
lattice\cite{BS04}
\begin{equation}
i\psi_t + \nabla_d^{2}\psi + \varepsilon \left[\sum_{i=1}^d \cos(2
x_i)\right]\psi + \gamma|\psi|^{2}\psi=0, \label{GPE}
\end{equation}
where $\nabla_d^2$ denotes the d-dimensional laplacian,
$\sum_{i=1}^d \cos(2 x_i)$ denotes a square optical lattice with
strength $\varepsilon$, $\gamma$ is the coefficient of
nonlinearity, and $x_i=x,y,z$ for $i=1,2,3$, respectively. Here we
will be mainly interested in cases $d=2$ and $d=3$. The existence
of localized solutions of the multidimensional GPE with periodic
potential and positive and negative nonlinearities (atomic
scattering lengths), has been previously investigated both by
variational analysis and by direct numerical simulations. In the
following we concentrate on a topic which was not discussed in
previous works, namely the existence of gap-Townes solitons in the
multidimensional GPE and its link to the phenomenon of
delocalizing transition. Due to the instability properties of
these solutions it is difficult to find them without an analytical
guide. To this regard we take advantage of the fact that the
periodic potential is separable and in spite of the nonlinearity
of the system we look for factorized stationary solutions of the
form
\begin{equation}
\psi(x_1,...,x_d)= \prod_{i=1}^d \phi_i(x_i)  e^{- i \mu t}.
\label{factorized}
\end{equation}
In 2D case ($d=2$) the substitution of the factorized ansatz into
Eq. (\ref{GPE}) gives:
\begin{equation}
\frac{\phi_{1_{xx}}}{\phi_1}+\frac{\phi_{2_{yy}}}{\phi_2}+\varepsilon
[\cos(2x)+\cos(2y)] + \gamma |\phi_1|^2|\phi_2|^2=-\mu.
\end{equation}
This equation  can also be written as
\begin{eqnarray}
\phi_{1_{xx}}+\varepsilon \cos(2x) \phi_1 + \frac\gamma 2
|\phi_2|^2 |\phi_1|^2 \phi_1 &=& - \mu_1 \phi_1, \nonumber \\
\phi_{2_{yy}}+\varepsilon \cos(2y) \phi_2 + \frac\gamma 2
|\phi_1|^2 |\phi_2|^2 \phi_2 &=& - \mu_2 \phi_2, \label{separable}
\end{eqnarray}
with $\mu_1=\mu_2=\mu/2$. By assuming $\phi_1=\phi_2\equiv \phi$
and adopting a diagonal coordinate $x=y\equiv \xi$ we have that
Eqs. (\ref{separable}) become equivalent to the following 1D
eigenvalue problem
\begin{equation}
\phi_{\xi \xi} + \varepsilon \cos (2 \xi) \phi + \frac \gamma 2
|\phi|^4 \phi= -\frac \mu 2 \phi. \label{quintic}
\end{equation}
From this we see that there is a link between the 2D cubic NLS and
the 1D quintic NLS which implies a rescaling of parameters as:
$\gamma \rightarrow \frac \gamma 2, \;\; \mu \rightarrow \frac \mu
2$.

The above equations can be easily extended  to the 3D GPE with
periodic potential. In this case Eq. (\ref{quintic}) will be
replaced by the following 1D NLS equation with septic nonlinearity
\begin{equation}
\phi_{\xi \xi} + \varepsilon \cos (2 \xi) \phi + \frac \gamma 3
|\phi|^6 \phi= -\frac \mu 3 \phi,  \label{septic}
\end{equation}
from which we see that in this case parameters must be rescaled
according to:  $\gamma \rightarrow \frac \gamma 3, \;\; \mu
\rightarrow \frac \mu 3$.

It is appropriate to mention that a factorized solution of the
form (\ref{factorized}) with the components solutions of the
nonlinear eigenvalue problem (\ref{quintic}) cannot be  an exact
solution of the 2D or 3D GPE,  since, due to the nonlinearity, the
problem is obviously not exactly separable. On the other hand, by
imposing the coincidence of the solutions along the diagonal axis
may be a constraint for a reasonable approximate solutions of the
2D and 3D problems, especially when the nonlinearity is small.
\begin{figure}\centerline{
\includegraphics[width=6.cm,height=6.cm,angle=0,clip]{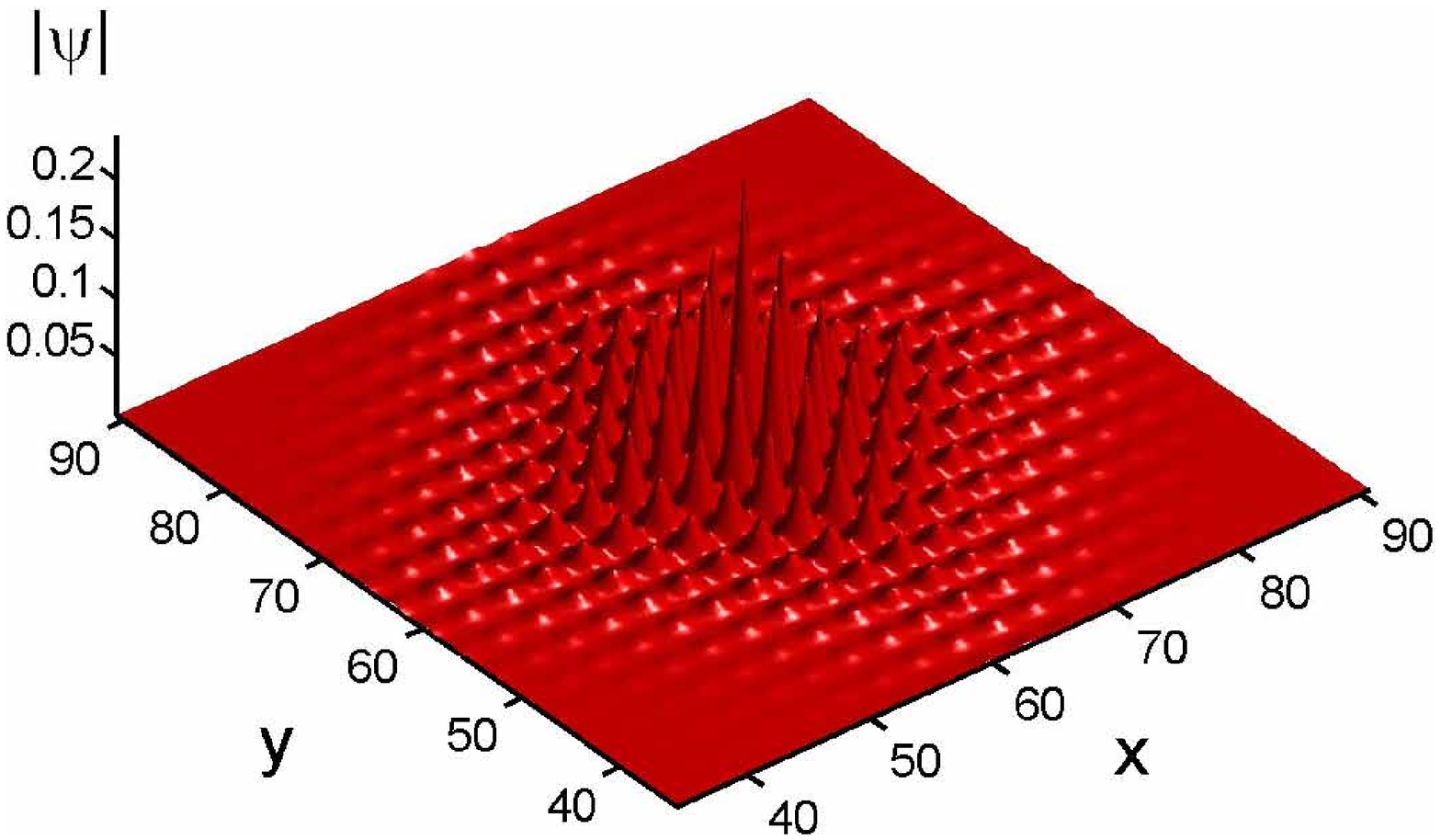}}
\vskip -.2cm \centerline{
\includegraphics[width=4.6cm,height=4.6cm,angle=0,clip]{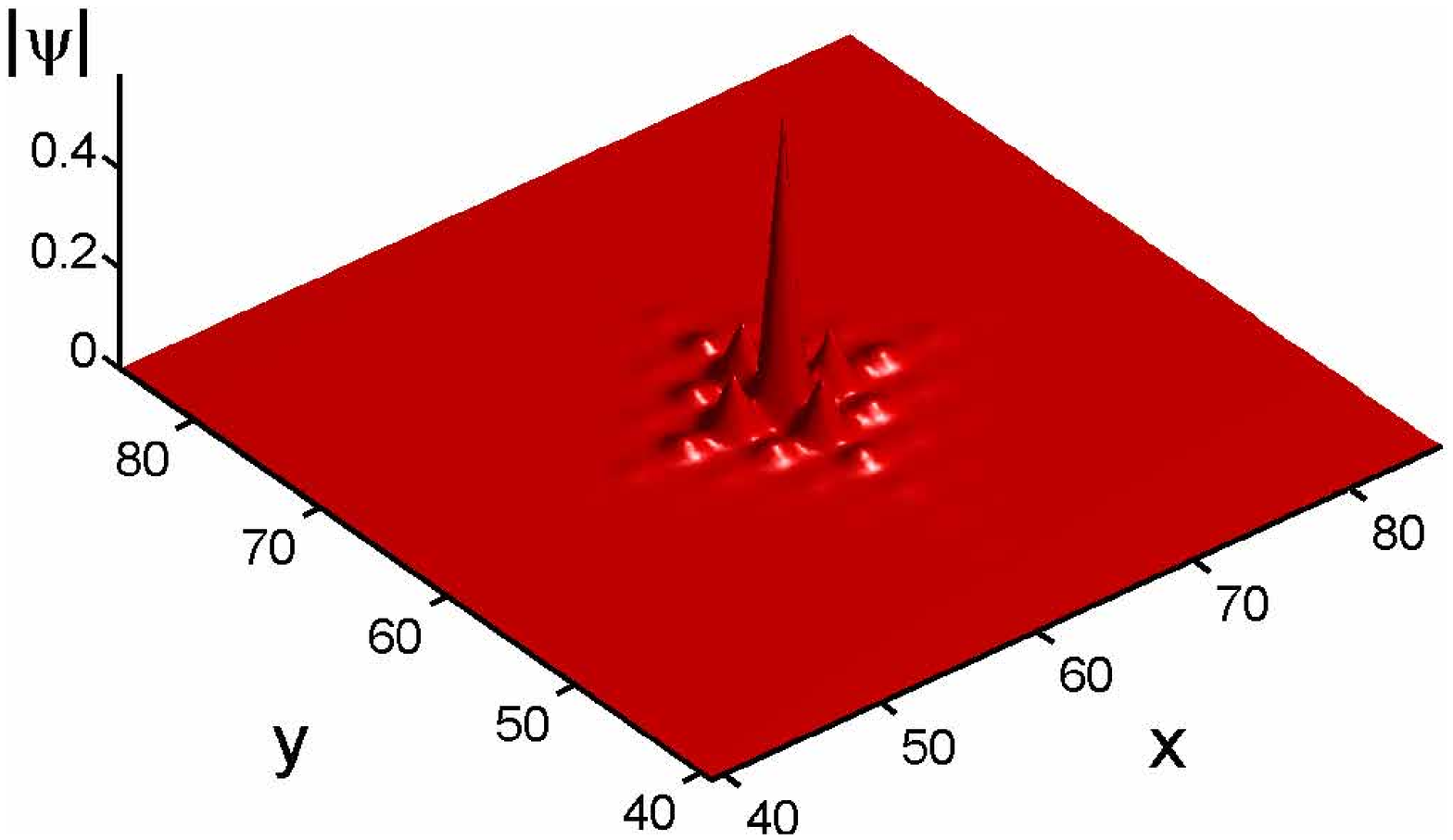}
\includegraphics[width=4.6cm,height=4.6cm,angle=0,clip]{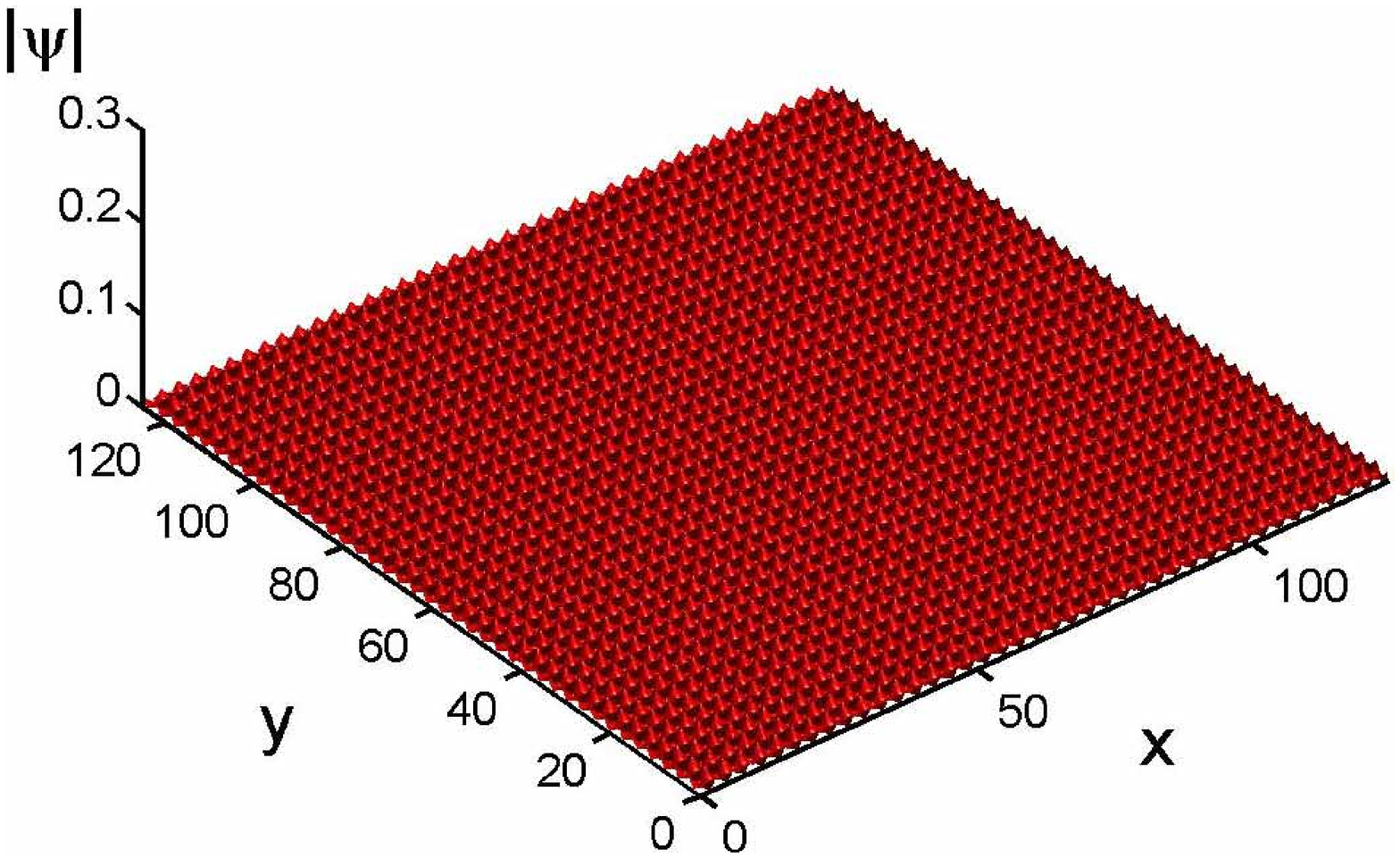}
} \caption{Gap-Townes soliton (top panel), gap soliton (lower left
panel)  and Bloch state (lower right panel) of the GPE for
parameter values $\varepsilon=5$, $\gamma=1$. The critical number
of atoms in normalized units for the gap-Townes soliton is
$N_c=0.4261$. The gap soliton and the Bloch state in the lower
panel are obtained for slightly overcritical (N=0.4347) and
undercritical (N=0.4176) values, respectively.} \label{fig1}
\end{figure}
In analogy with the 1D NLS with quintic nonlinearity investigated
in Ref. \cite{AS05} we expect that the delocalizing curve
coincides with the existence curve of gap-Townes solitons for
which it was shown that the critical number of atoms decreases
with increasing the strength of the OL. This means that in a deep
OL the effective nonlinearity required for the existence of a
gap-Townes soliton is smaller and the problem may become
effectively close to separable.

To check the correctness of this argument we construct factorized
solutions (\ref{factorized}) of the 2D GPE  by means of a
self-consistent method which allows to solve  the 1D quintic NLS
eigenvalue problem exactly (similar results can be obtained for
the 3D case).

In the top panel of Fig. 1 we show a 2D gap-Townes soliton
obtained from the product ansatz using exact (self-consistent)
gap-Townes solitons of the corresponding quintic NLS equation in
Eq. (\ref{quintic}). The lower left and right panels show,
respectively, the gap soliton and the extended Bloch state found
at energy slightly below and slightly above (bottom of the lowest
band) the one of the gap-Townes soliton. To check the reliability
of the factorized ansatz we have computed the time evolution under
the original 2D GPE equation using the factorized solution as
initial condition. This is shown in Fig. 2 where the time
evolution of the gap-Townes soliton in Fig. 1 (central panel) and
the ones obtained  for slightly overcritical and undercritical
numbers of atoms are shown.
\begin{figure}\centerline{
\includegraphics[width=3.5cm,height=6.7cm,angle=0,clip]{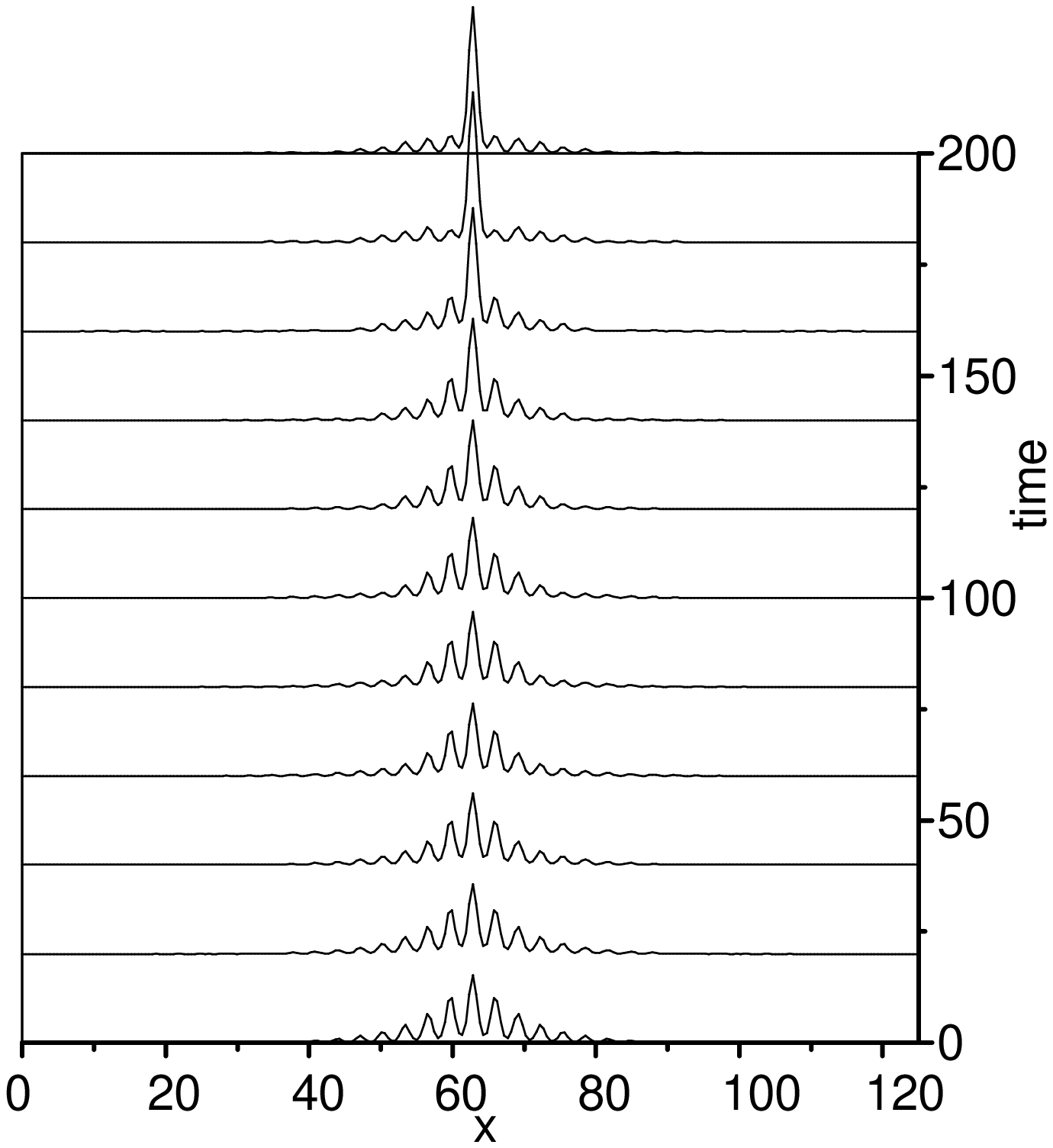}
\includegraphics[width=3.5cm,height=6.2cm,angle=0,clip]{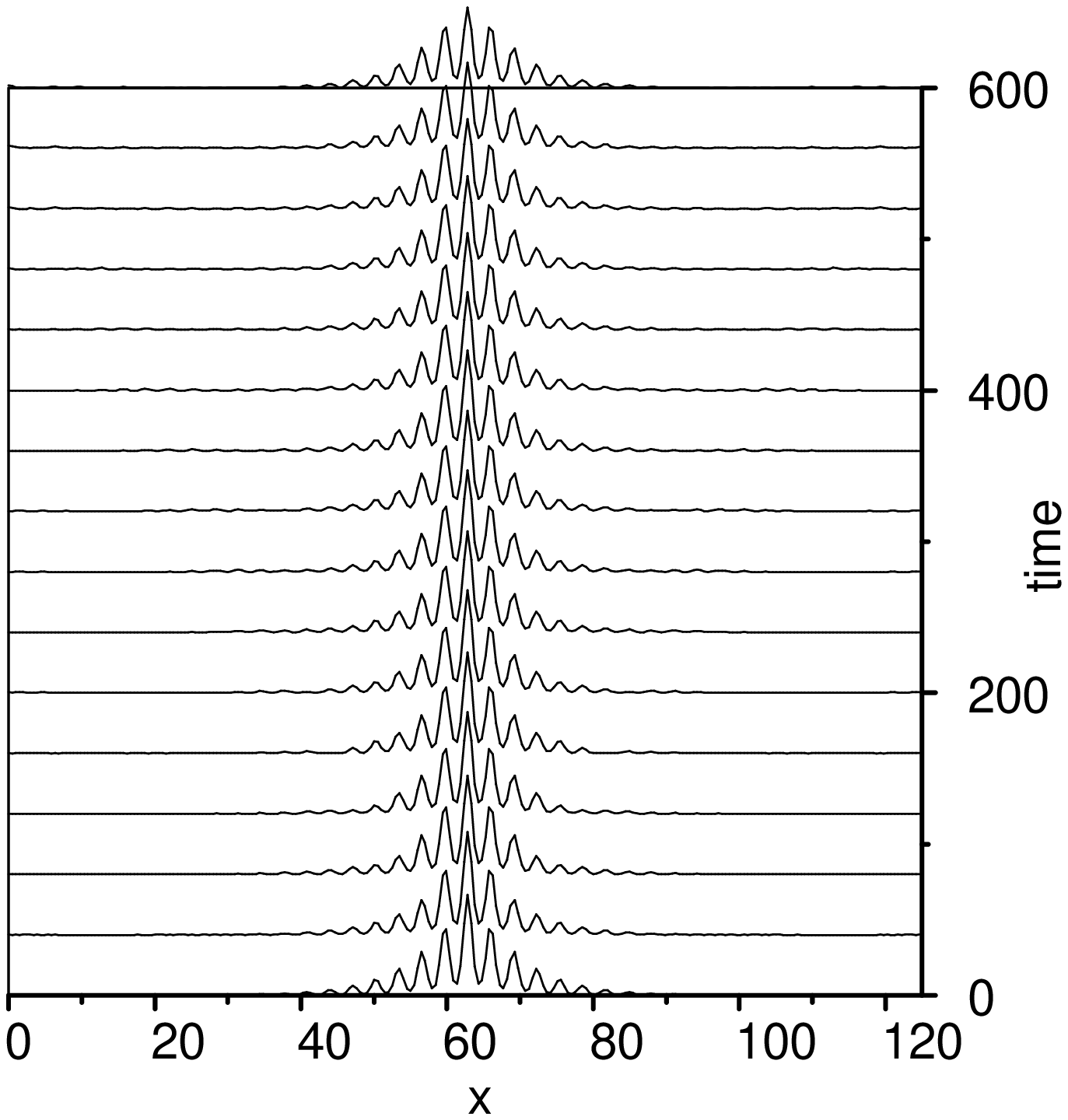}
\includegraphics[width=3.5cm,height=5.8cm,angle=0,clip]{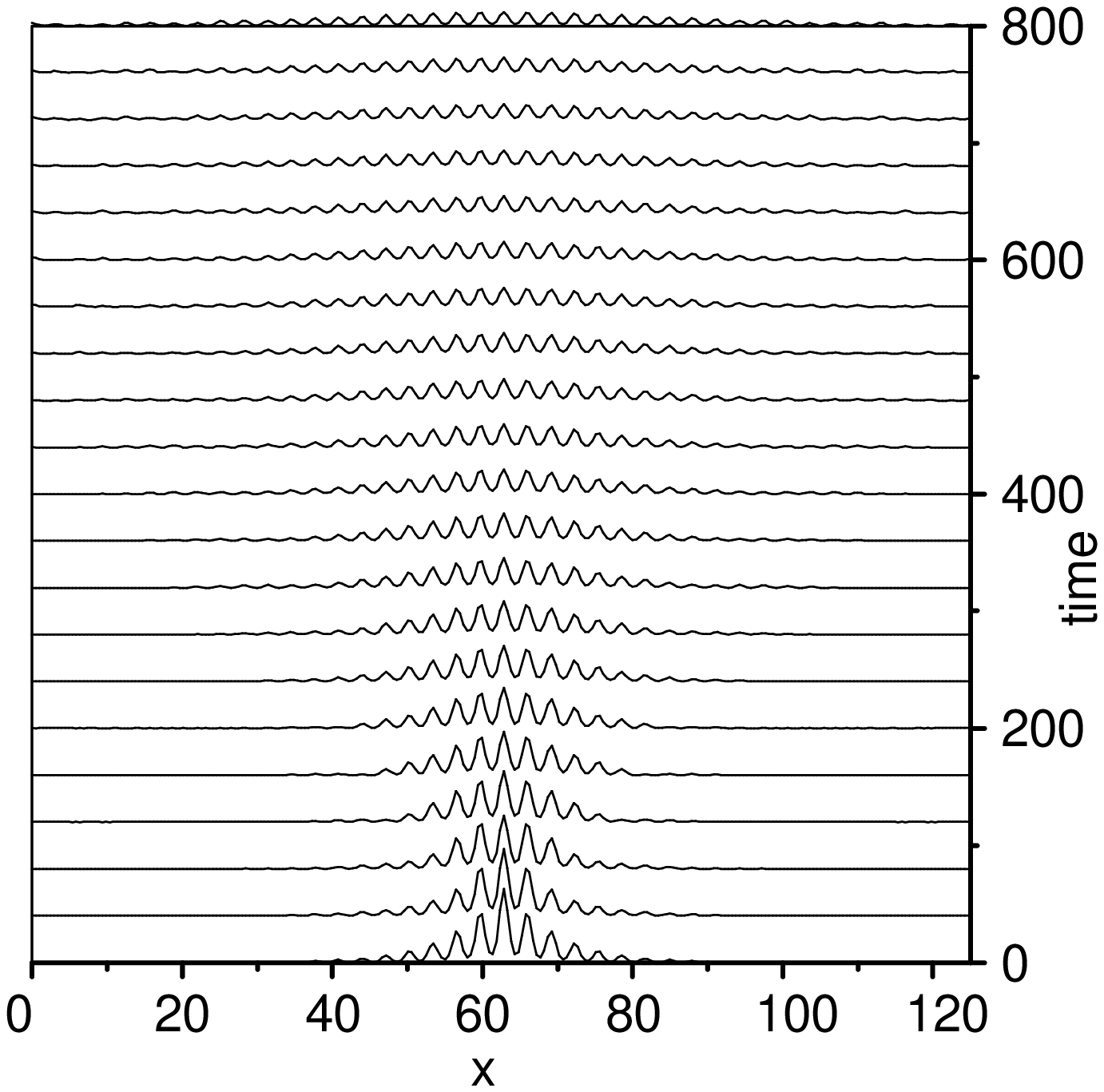}
} \caption{Time evolution of the gap-Townes soliton shown in Fig.
1 with critical number of atoms  (central panel) and for slightly
overcritical (left panel) and undercritical (right panel) number
of atoms.
 } \label{fig2}
\end{figure}
We see that while the product solution constructed from the
quintic NLS remains localized for a long time, a slight increase
or decrease of the number of atoms produces shrinking or decay of
the solution, respectively. This clearly shows the existence and
role of gap-Townes solitons of the 2D GPE with optical lattice in
characterizing the delocalizing threshold.

\section{Variational analysis and  existence of localized states}

\begin{figure}\centerline{
\includegraphics[width=3.9cm,height=3.9cm,angle=0,clip]{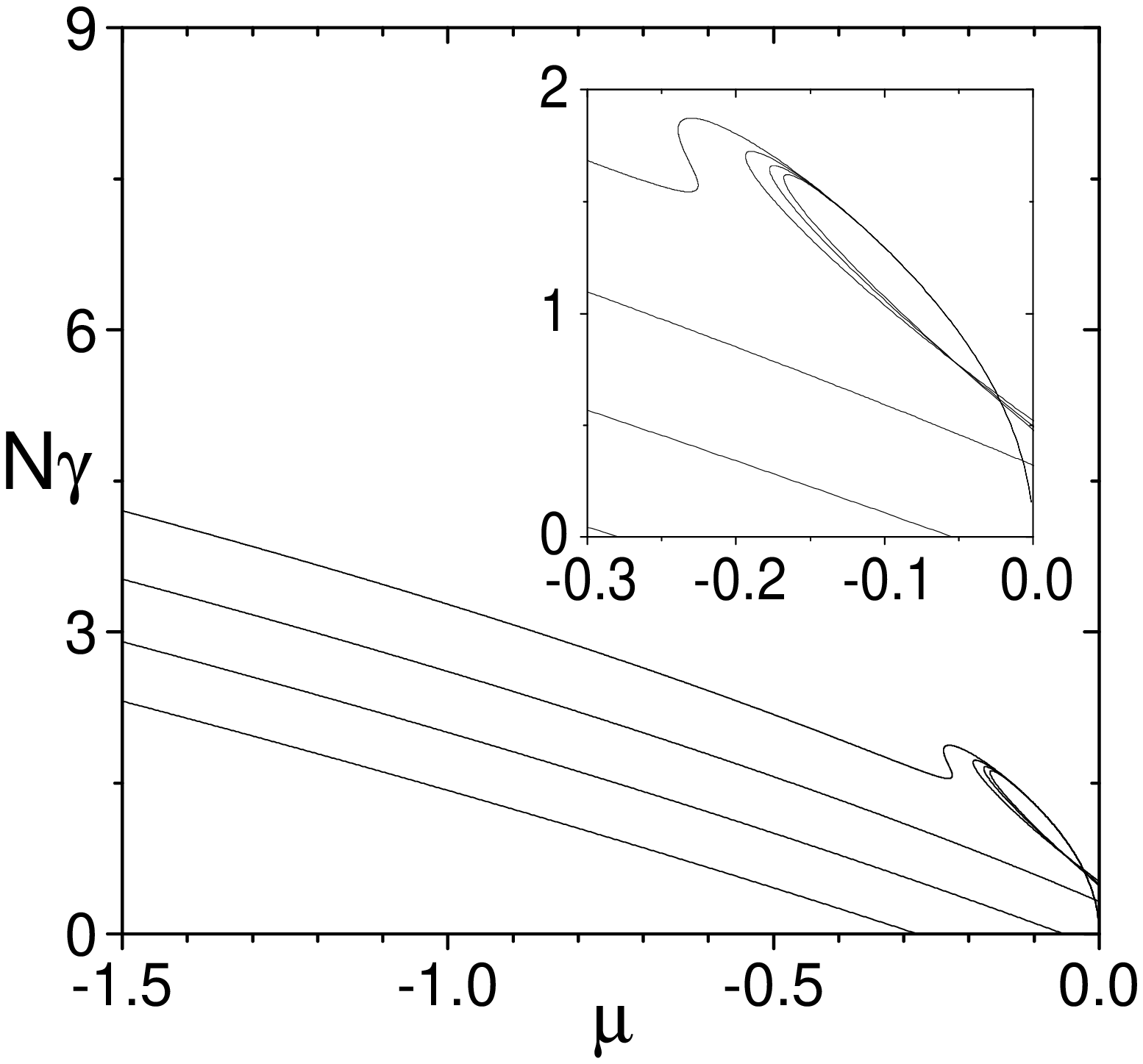}
\includegraphics[width=3.9cm,height=3.9cm,angle=0,clip]{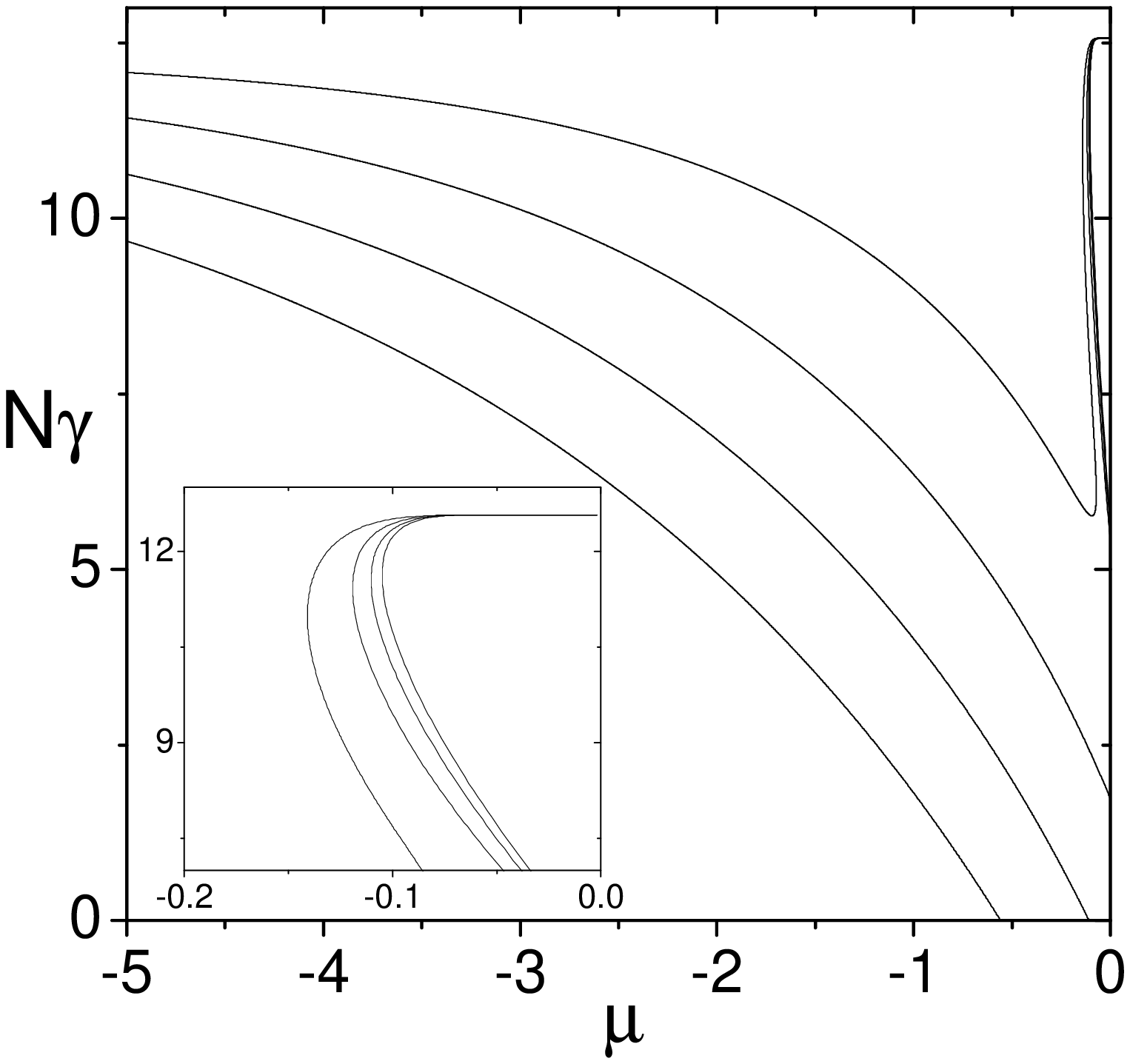}
\includegraphics[width=3.9cm,height=3.9cm,angle=0,clip]{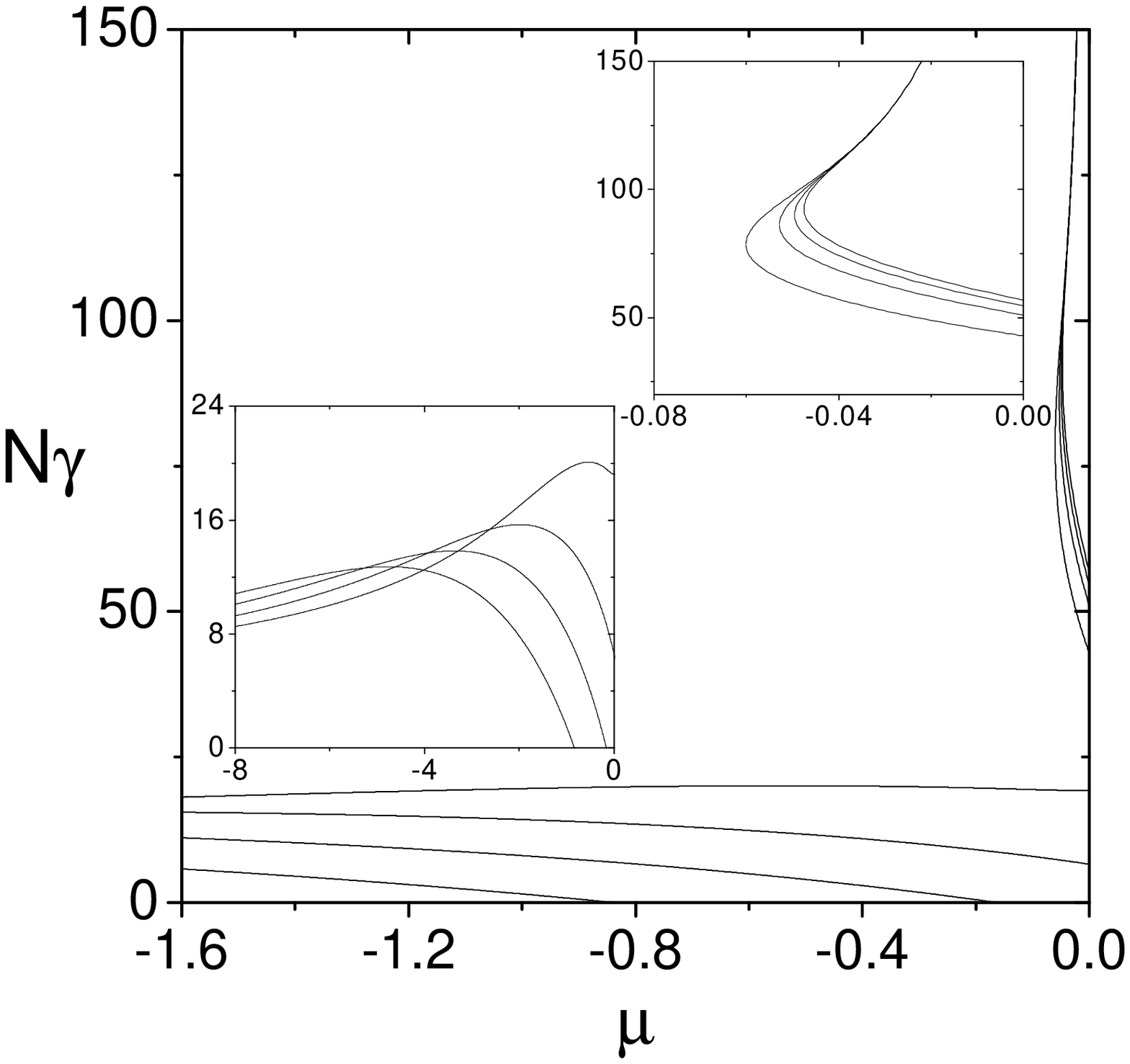}}
\caption{Left panel. VA existence curves for the 1D case ($d=1$ in
Eqs. (\ref{vaeq})) for different values of $\varepsilon$. The
inset shows an enlargement for small values of $\mu$. Curves refer
to $\varepsilon$ values increased in steps of $0.5$ starting from
$\varepsilon=0.5$ (upper curve and most left curve in the inset)
to $\varepsilon=2.0$ (lower curve and most right curve in the
inset). Central panel. Same as for the left panel but for the 2D
case ($d=2$ in Eqs. (\ref{vaeq})). The inset shows an enlargement
for small values of $\mu$ close to the threshold $N=4 \pi$. Right
panel. Same as for the left panel but for the 3D case ($d=3$ in
Eqs. (\ref{vaeq})). The upper inset shows an enlargement of the
curves close to the $\mu=0$ axis, while the lower inset displays
the curves close to the $N\gamma=0$ axis, on a larger scale.}
\label{fig3}
\end{figure}

The existence of localized states in multidimensional GPE with OL
and in 1D NLS with higher order nonlinearities can be investigated
by means of the variational approach (VA). To this regard we first
consider Eq. (\ref{GPE}) and search for stationary solutions of
the form $\psi = Ue^{-i\mu t}$ and consider the Lagrangian density
associated with the equation for the stationary field $U$
\begin{eqnarray}
{\cal L}= \frac{1}{2}(\nabla_d U)^2 - \frac{\mu}{2}U^2 -
\frac{\varepsilon}{2}\left[\sum_{i=1}^d \cos(2 x_i)\right] U^2  -
\frac{\gamma}{4}U^{4}.
\end{eqnarray}
By taking a Gaussian ansatz for $U$
\begin{equation}
U = Ae^{-\frac{a}{2}\sum_{i=1}^d x_i^2},
\end{equation}
and performing spatial integration we obtain the following
effective lagrangian $L_{eff}=\int {\cal L} dx$ for parameters $A,
a$
\begin{equation}
L_{eff} = \frac {A^2}{2} \left(\frac \pi a\right)^{\frac d 2}
\left[\frac {d}{2} a- \mu - d \varepsilon e^{-1/a}- \frac{A^2
\gamma}{2^{\frac{d}{2}+1}} \right], \;\; d=1,2,3.
\end{equation}
Variational parameters $A, a$, indicating the amplitude and
inverse width of the localized state, are linked to the number of
atoms by the relation $N = A^{2} (\frac{\pi}{a})^{d/2}$. From the
conditions of stationarity of the effective lagrangian $\partial
L_{eff}/\partial a=\partial L_{eff}/\partial A=0$,  we get the
following equations relating $N$ with $a$ and the chemical
potential $\mu$
\begin{eqnarray}
\label{vaeq} \mu &=& \frac{d}{2}a - d\varepsilon e^{-1/a} -
\frac{\gamma N}{2^{d/2}}
\left(\frac a \pi\right)^{d/2},\\
N &=& \frac{ 4 \pi^{d/2}}{\gamma} \left(\frac 2 a
\right)^{\frac{d}{2}-1} \left(1-\frac{2
\varepsilon}{a^2}e^{-1/a}\right), \qquad d=1,2,3. \nonumber
\end{eqnarray}
In Fig. \ref{fig3} we depict the ($N, \mu$) curves obtained from
the above transcendental equations for fixed values of
$\varepsilon$ and for the cases $d=1,2,3$. We see that for the 1D
case $dN/d\mu$ is always negative, this means, according to the
Vakhitov-Kolokolov (V-K) criterion \cite{Vakhitov}, that the
solution exists and is stable for any value of $\varepsilon$
without limitations on the effective nonlinearity $N \gamma$. In
the 2D case, a threshold in $N \gamma$ appears which is  predicted
by VA to be  exactly $4\pi$ for $\varepsilon = 0$. Also notice
that in this case $dN/d\mu <0$ is still satisfied for most branch
curves, meaning that the solution is usually stable. The situation
is quite different in the 3D case, where there is no limiting
threshold for existence but most of the curves display a positive
slope meaning that the solution is unstable. In particular, from
the upper inset of the right panel we see that for curves close to
the $\mu=0$ axis, $dN/d\mu$ change signs at $N \gamma\approx 70$
and becomes positive for higher values of $N\gamma$. The almost
horizontal curves for lower values of $N\gamma$ are displayed in
the lower inset of the figure,  from which we see that $dN/d\mu$
changes from negative to positive after the curves have reached a
maximum at values of $\mu$ which depend on $\varepsilon$.

Notice that for the 3D case the above equations predict for
$\varepsilon = 0$ the number of atoms dependence on the chemical
potential as
$$
N = \frac{4\pi\sqrt{\pi}}{\gamma\sqrt{-\mu}}.
$$
Similar dependence $N$ vs. $\mu$ was previously obtained for
stationary solutions of 3D NLS with cubic and quintic nonlinearity
\cite{Akhmediev}. Also notice that according to V-K criterion this
solution is unstable. From the condition $N \geq 0$,  we obtain
the limitation on the soliton width as
\begin{equation}
\frac{2\varepsilon}{a^2}e^{-1/a} \leq 1,
\end{equation}
while for $\varepsilon >> 1$ we obtain that the soliton exists if
the width satisfies the condition
\begin{equation}
a < a_{c1} \approx \frac{1}{2(\log(\delta) + \log(\log(\delta)))},
\ \delta = 2\sqrt{2\varepsilon},
\end{equation}
and $a > a_{c2} \approx  \sqrt{2\varepsilon}-1/2$. For
$\varepsilon = 5$, for example, we obtain $a_{c1} = 0.2$, while
the exact value is $a_{e} \approx 0.17$.

In view of the analogy of the 2D and 3D GPE with the 1D NLS with
quintic and septic nonlinearity, respectively, it is of interest
to compare the above VA equations with these cases. To this
regard, we consider the 1D GPE with a high order nonlinearity of
the type $\gamma |\psi|^\alpha \psi$ with $\alpha=2,4,6$
\begin{equation}
i\psi_t + \psi_{xx} + \gamma |\psi|^\alpha \psi + \varepsilon
\cos(2 x) \psi = 0. \label{q-sept}
\end{equation}

Using the same approach as before, one can readily show that the
effective lagrangian in this case is
\begin{equation}
L_{eff} = \frac {A^2}{2} \left(\frac \pi a\right)^{1/2}
\left[\frac {1}{2} a- \mu - \varepsilon e^{-1/a}- \frac{A^\alpha
\gamma}{\left(\frac{\alpha+2}{2}\right)^{3/2}} \right],
\end{equation}
from which the following VA equations are derived
\begin{eqnarray}
\label{vaeq} \mu &=& \frac{a}{2} - \varepsilon e^{-1/a} -
\frac{\gamma N^{\alpha/2}}{(\frac{\alpha+2}{2})^{1/2}} \left(\frac
a \pi \right)^{\alpha/4} ,\\
N^{\alpha/2} &=& {\frac{2 a}{\alpha \gamma}}
\left(\frac{\alpha+2}{2}\right)^{3/2} \left(\frac \pi
a\right)^{\alpha/4} \left(1-\frac{2
\varepsilon}{a^2}e^{-1/a}\right), \nonumber
\end{eqnarray}
here $\alpha=2,4,6$. We see that the case $\alpha=2$ coincides
with the case $d=1$ considered above, and the case $\alpha=4$ with
the quintic VA equations derived in \cite{AS05}.

Notice that for $\varepsilon=0$ Eq. (\ref{q-sept}) admits exact
solutions also for $\alpha=4,6$. For the septic case, indeed, we
have, using $\psi = ue^{imt}, m = -\mu > 0$, that
\begin{equation}
\psi = (\frac{4m}{\gamma})^{1/6}\mbox{sech}^{1/3}(3\sqrt{m}x),
\end{equation}
is an exact  solution with a norm
\begin{equation}
N = \frac{2^{1/3}\Gamma^{2}(\frac 1 3)} {3\Gamma(\frac 2
3)\gamma^{1/3} m^{1/6}},
\end{equation}
where $\Gamma(x)$ is the gamma function. The Hamiltonian for this
solution is equal to
$$
H = \int_{-\infty}^{\infty}(|\psi_{x}|^2 -
\frac{\gamma}{4}|\psi|^8 )dx = 0.44 \frac{m^{5/6}}{\gamma^{1/3}}.
$$
From the above VA equations for the 2D case, one can derive the
value of $N_c$ for small values of $\varepsilon \neq 0$ as: $N_c =
4\pi(1-8\varepsilon\exp(-2)).$ For  $\varepsilon = 0.2$ we obtain
$N_c=9.845$ which is in reasonable agreement with the value $10.8$
obtained from numerical simulations of the 2D case. We remark that
for small values of $\varepsilon$ the soliton is very extended in
space and resembles a Bloch wave modulated with an envelope. In
this case an effective mass approximation may be appropriate which
allows to replace the GPE field equation with a nonlinear
Schr\"odinger equation with effective mass $ m^{\ast}$ and
nonlinearity $\beta$. In the 2D case we have
\begin{equation}
iu_{t} + \alpha (u_{xx}+u_{yy}) + \beta |u|^{2}u =0,
\end{equation}
where $\alpha = m^{\ast}/m,  m^{\ast(-1)} = (\partial^2 E/\partial
k^2)_ {k=0}$ and $\beta = (2\pi/L^2)\int d^2 r |\phi_{1,0}|^4$ (a
similar equation can be written also for the 3D case) . In this
approximation the norm of the gap Townes soliton can be evaluated
as $N = \frac{N_{T}}{\alpha\beta}$. For deep optical lattices
($\varepsilon > 5$) $\alpha$ can be approximated as  $\alpha
\approx \varepsilon^{1/4}$ and the norm $N \approx
\frac{N_{T}}{\varepsilon^{1/4} \beta}$.

\section{Numerical study of gap-Townes solitons and delocalizing
transitions}

In this section we investigate  the existence curve of gap-Townes
solitons in the $(N, \varepsilon)$ plane. As mentioned before,
this curve coincides with  the delocalizing transition curve which
separates stable localized solutions from decaying ones. Its
knowledge is therefore important for experimental investigations
of multidimensional solitons. To this regard we remark that for
the observation of multidimensional BEC solitons parameters should
be chosen between the delocalizing and the collapsing curves. In
the following we investigate the delocalizing curve by means of
direct numerical integrations of the 2D and 3D GPE with periodic
potential and by the corresponding 1D NLS systems with quintic and
septic nonlinearities discussed above, respectively. The existence
of gap-Townes solitons is then shown by direct numerical
simulations of the multidimensional GPE using as initial
conditions the above mentioned product states, which are found by
solving the 1D GPE with higher nonlinearities by means of the
self-consistent method described in \cite{ms05}.
\begin{figure}\centerline{
\includegraphics[width=4.2cm,height=4.2cm,angle=0,clip]{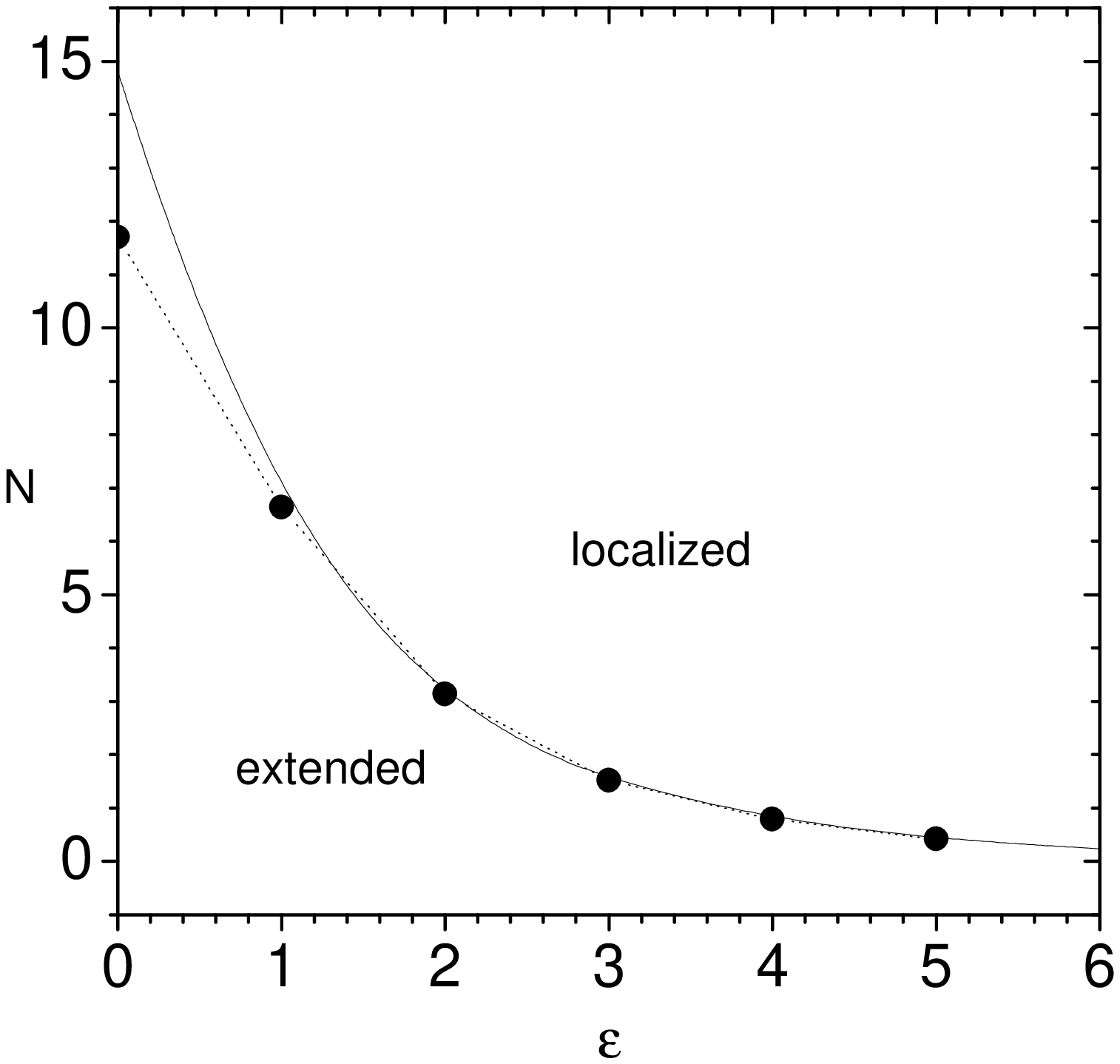}
\includegraphics[width=4.2cm,height=4.2cm,angle=0,clip]{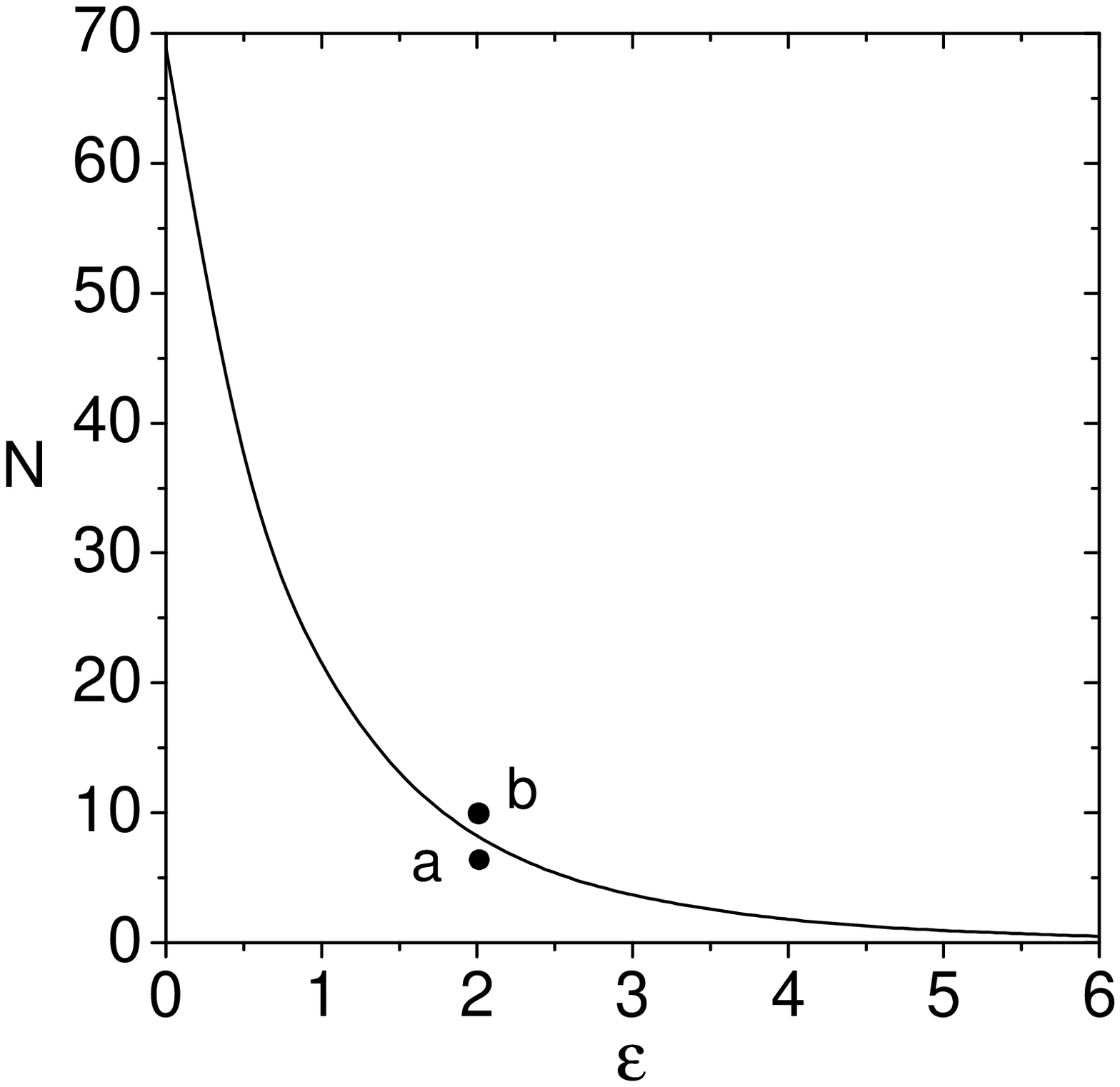}}
\caption{Left panel: Existence curve of gap-Townes solitons of the
2D GPE for $\gamma=1$. The continuous line represents the
delocalizing curve as obtained from the 1D quintic NLS
approximation. The dots joined by thin dotted line represent
numerical results obtained from direct integrations of the 2D GPE.
Right panel: Existence curve of gap-Townes solitons of the 3D GPE
for $\gamma=1$. The line represents the delocalizing curve as
obtained from the 1D quintic NLS approximation, while the dots
represent numerical results obtained from direct integrations of
the 3D GPE.} \label{fig4}
\end{figure}
In Fig. \ref{fig4} we depict the existence curve of gap-Townes
solitons of the 2D GPE for $\gamma=1$ as obtained from numerical
integrations of the 2D GPE. The corresponding curve obtained from
the 1D quintic NLS approximation by means of a self-consistent
approach is also shown. We see that for $\varepsilon > 1$ the 1D
quintic NLS curve agrees very well with that of the 2D GPE, the
deviations becoming evident only for strengths of the OL which are
less than $\varepsilon \approx 1$ (one recoil energy). This fact
can be easily understood from the observation that for a fixed
value of $\varepsilon$ there is one value of $N$ for which the gap
Townes soliton exists and that by increasing $\varepsilon$ the
corresponding value of N decreases. This implies that for
gap-Townes solitons in a strong OL the nonlinear interaction is
effectively small,  due to the potential barriers which prevent
tunneling of  matter into adjacent wells. On the contrary, in a
shallow optical lattice the matter can easily tunnel through the
barriers and the effective (attractive) nonlinearity can be
larger. Since the separability ansatz used to link the 2D GPE to
the 1D quintic NLS works well when the nonlinearity is small, it
is clear that a discrepancy can arise at small values of
$\varepsilon$. The fact that the quintic NLS equation deviates
from the 2D GPE only for $\varepsilon <1$, however, makes the
mapping between these two equations very convenient for practical
calculations.
\begin{figure}
\centerline{
\includegraphics[width=4.1cm,height=4.9cm,angle=0,clip]{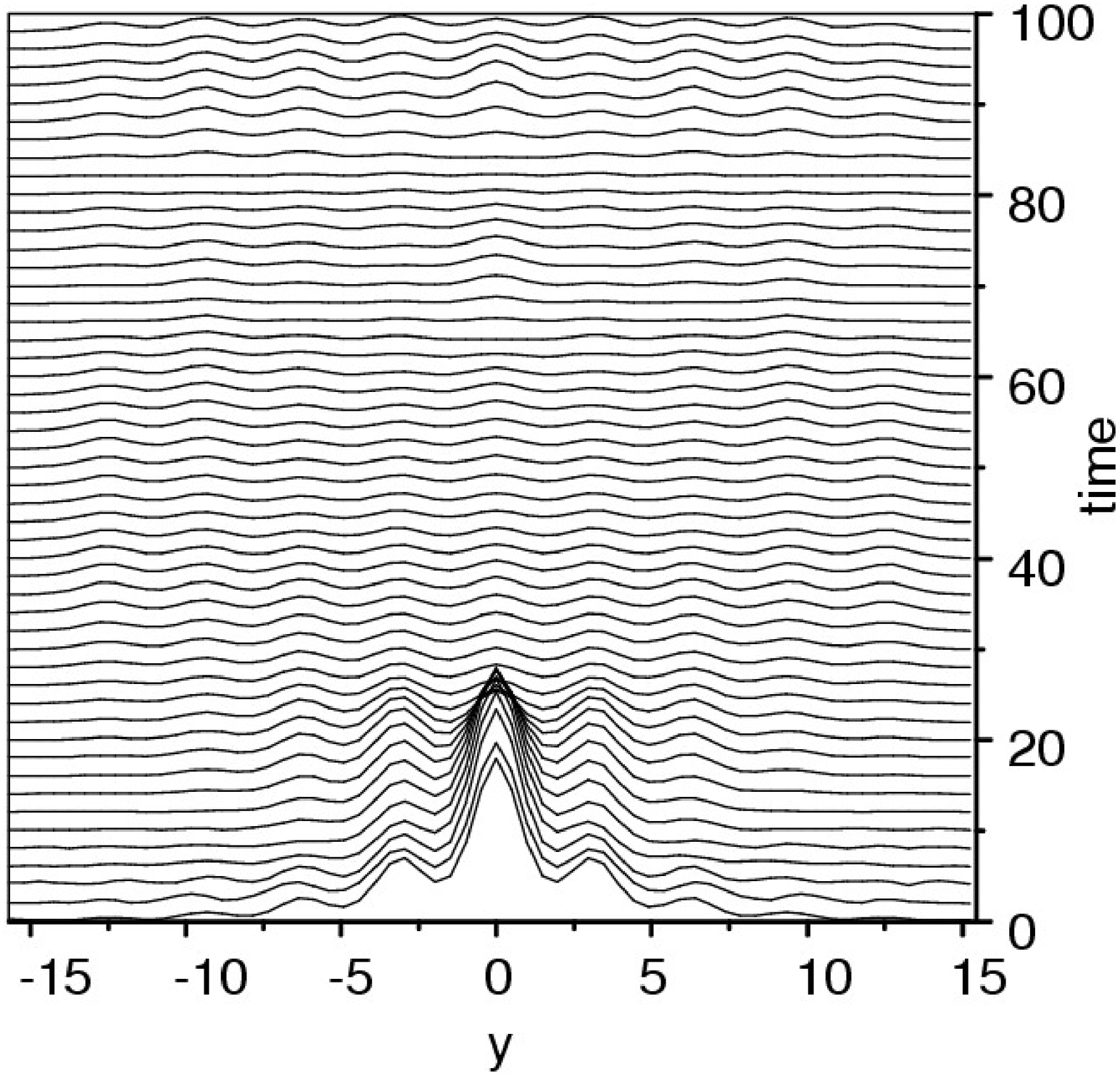}
\includegraphics[width=4.1cm,height=6.2cm,angle=0,clip]{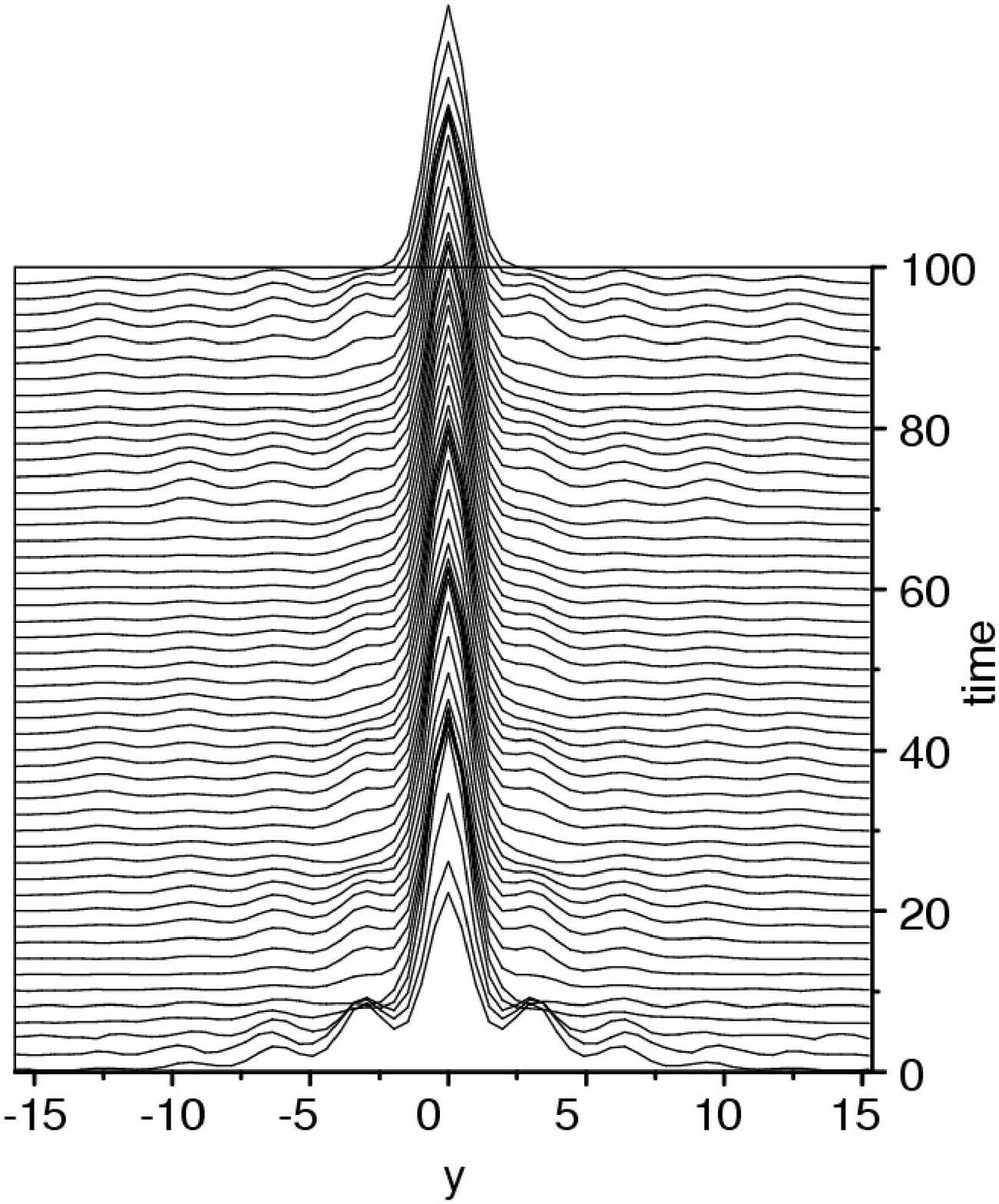}
} \caption{Time evolution of the y-section of a 3D gap-Townes
soliton obtained for  $\varepsilon=2$ with slightly undercritical
(left panel) and overcritical (right panel) number of atoms,
corresponding to points a and b in Fig \ref{fig4}, respectively.}
\label{fig5}
\end{figure}
A similar situation seems to be true also for the 3D case. In the
right panel of Fig. \ref{fig4} we depict the existence curve of
gap-Townes solitons of the 3D GPE for $\gamma=1$ as obtained from
the 1D septic NLS approximation by means of a self-consistent
approach. Due to the long computational times required in 3D
simulations we have presented verification for only few points of
the curve. A more complete analysis will require further
investigation. In particular, the prediction of our analysis are
checked by means of numerical integrations of the 3D GPE in
correspondence of the two filled dots depicted in the right panel
of Fig. \ref{fig4} at $\varepsilon=2$, just above and below the
delocalizing curve obtained from the septic 1D NLS equation. In
Fig. \ref{fig5} we show the time evolution of the y-section of a
3D gap-Townes soliton obtained for $\varepsilon=2$ in
correspondence of these slightly undercritical (left panel) and
overcritical (right panel) points. We see that, in analogy to what
we observed in 2D case, the initial condition with an
undercritical number of atoms leads to the complete delocalization
of the state, while in the overcritical case a gap soliton state
which remains stable over long time is formed. It is appropriate
to mention that the exact profile of the 3D gap-Townes soliton,
which separates these two behaviors is more difficult to obtain
with the separability ansatz than for the previous 2D case, since
in the 3D case this ansatz appears to be less accurate. The
signature of the gap-Townes soliton state, however, is very clear
as one can see from the early stages of the time evolution
depicted in Fig. \ref{fig5} (notice from the left panel that the
undercritical state remains stable up to a time $t=20$ before
starting to decay).  This behavior strongly suggests the existence
of gap-Townes solitons also in the 3D case.

\begin{figure}
\centerline{
\includegraphics[width=12cm,height=6.cm,clip]{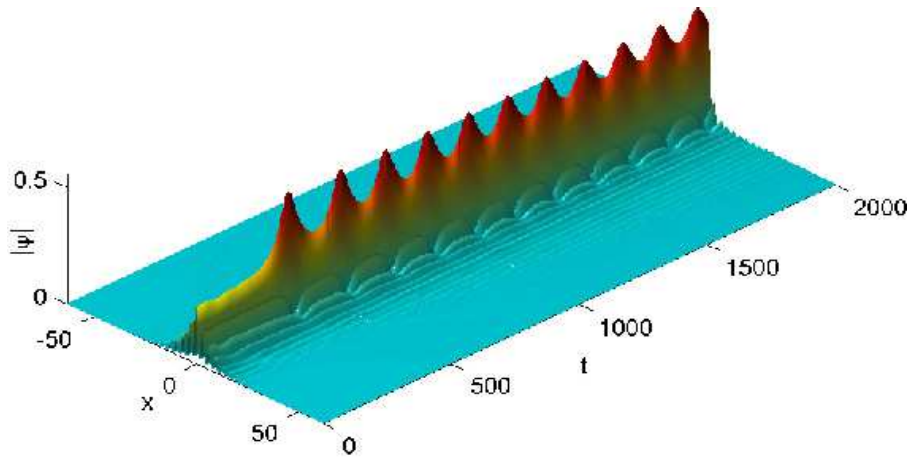}}
\caption{Oscillations of the gap-Townes soliton when it collapses
into the final gap state, according to Eq. (\ref{GPE}) with
initial state taken as the product solution constructed from two
quintic NLS with a total norm increased by one percent. Shown is
the $y=0$ cross section of the 2D profile.} \label{fig6}
\end{figure}

We now address to another important property of the gap-Townes
soliton which shows up in the transition to a gap soliton state
for overcritical number of atoms. As it was already mentioned,
even a slightly overcritical norm gives rise to a rapid shrinking
of the gap-Townes soliton. Although at initial stage of evolution
this behavior is similar to the collapse of ordinary Townes
soliton, the final state is different. Specifically, gap-Townes
soliton approaches the (final) gap state via long-lasting
oscillations, as shown in Fig. \ref{fig6}. Each "reflection" from
the broad state is accompanied by emission of linear waves, which
can be viewed as a tunneling of matter from the localized mode
into the extended one. This process also contributes to the
damping of the oscillations. We remark that these oscillations,
having a very regular behavior, could be used to detect the
existence of gap-Town solitons in a real experiment.

\section{Conclusions}
In this paper we have shown the existence of gap-Townes solitons
in the multidimensional (2D and 3D) Gross-Pitaevskii equation with
a periodic potential and discussed its link with the phenomenon of
delocalizing transition. These solutions have the peculiarity of
being unstable under small fluctuations of the number of atoms and
separate localized (soliton like) states from extended (Bloch
like) ones. The existence curve in the parameter space of this
particular solution is very useful since it provides the lower
threshold for the existence of localized states. The gap-Townes
solitons discussed in this paper are a natural generalization of
the Townes solitons of the nonlinear Schr\"odinger equation
(without periodic potential). The existence of Townes solitons in
a nonlinear glass sample modelled by the NLS equation was
experimentally demonstrated in Ref. \cite {exp-townes}. The fact
that the transition from a gap-Townes soliton to a gap soliton is
always accompanied by regular oscillations gives the possibility
to indirectly observe the multidimensional gap-Townes solitons
discussed in this paper in real experiments. In particular we
expect these solitons to be observed both in multidimensional BECs
in OL and in nonlinear optics systems, including 2D and 3D
photonic crystals and arrays of nonlinear optical waveguides.

\section{Acknowledgements}
MS acknowledge partial financial support from a MURST-PRIN-2003
and a MIUR-PRIN-2005 initiatives. FKhA and BBB wish to thank the
Department of Physics "E. R. Caianiello" for hospitality and the
University of Salerno for financial support.


\begin{thebibliography}{99}

\bibitem{Burger}
S. Burger, K. Bongs, S. Dettmer, W. Ertmer, K. Sengstock, A.
Sanpera, G. V. Shlyapnikov, and M. Lewenstein, Phys. Rev. Lett.
{\bf 83}, 5198 (1999).

\bibitem{Khaykovich}
L. Khaykovich, F. Schreck, G. Ferrari, T. Bourdel, J. Cubizolles,
L. D. Carr, Y. Castin and C. Salomon, Science {\bf 296}, 1290
(2002).

\bibitem{Strecker}
K. E. Strecker, G. B. Partridge, A. G. Truscott and R. G. Hulet,
Nature {\bf 417}, 150 (2002).

\bibitem{Eiermann}
B. Eiermann, P. Treutlein, Th. Anker, M. Albiez, M. Taglieber,
K.-P. Marzlin, and M. K. Oberthaler, Phys. Rev. Lett. {\bf 91},
060402 (2003).

\bibitem{Sulem}
C. Sulem and P. L. Sulem, {\it The Nonlinear Schr\"odinger
Equation: Self-focusing and Wave Collapse}, Springer, 1999.

\bibitem{townes}
R. Y. Chiao, E. Garmire and C. H. Townes, Phys. Rev. Lett. {\bf
13}, 479 (1964).

\bibitem{BKS}
B. B. Baizakov, V. V. Konotop, and M. Salerno, J. Phys. B: At.
Mol. Opt. Phys., {\bf 35 }, 5105 (2002).

\bibitem{Efremidis02}
N. K. Efremidis, S. Sears, D. N. Christodoulides, J. W. Fleischer
and M. Segev, Phys. Rev. E {\bf 66}, 046602 (2002).

\bibitem{OK03}
E. Ostrovskaya and Yu.S. Kivshar, Phys. Rev. Lett. {\bf 90},
160407 (2003).

\bibitem{BMS03}
B. B. Baizakov,  B. A. Malomed, and M. Salerno,  Eurphys. Lett.
{\bf 63}, 642 (2003).

\bibitem{Lederer05} D. Mihalache, D. Mazilu, F. Lederer, B. A. Malomed, L.-C.
Crasovan, Y. V. Kartashov, and L. Torner, Phys. Rev. A {\bf 72},
021601(R) (2005).

\bibitem{BS04}
B. B. Baizakov  and  M. Salerno,  Phys.Rev. A {\bf 69}, 013602
(2004).

\bibitem{AS05}
F. Kh. Abdullaev and M. Salerno, Phys. Rev. A {\bf 72}, 033617
(2005).

\bibitem{KP}
G.L. Alfimov, V.V. Konotop, P. Pacciani, Phys. Rev. A {\bf 75},
023624 (2007).

\bibitem{CBKS08}
H.A. Cruz, V.A. Brazhnyi, V.V. Konotop, and M. Salerno, Physica D
(2008), doi:10.1016/j.physd.2008.09.008.

\bibitem{Berge}
L. Berge,  Phys. Rep.  {\bf 303}, 259 (1998).

\bibitem{Efremidis03}
N. K. Efremidis, J. Hudock, D. N. Christodoulides, J. W.
Fleischer, O. Cohen, and M. Segev, Phys. Rev. Lett. {\bf 91},
213906 (2003).

\bibitem{Vakhitov}
N. G. Vakhitov, A. A. Kolokolov, Radiophysics and Quantum
Electronics, {\bf 16}, 783 (1973).

\bibitem{Akhmediev} N. Akhmediev, M. P. Das and A. V. Vagov, Int.
J. Mod. Phys. B {\bf 13} 625 (1999).

\bibitem{ms05} Mario Salerno, Laser Physics {\bf 15}, No.4, 620
(2005).

\bibitem{exp-townes}
K. D. Moll, A. L. Gaeta, G. Fibich, Phys. Rev. Lett. {\bf 90},
203902 (2003).

\end{thebibliography}
\end{document}